\documentclass[twocolumn,12pt]{article}
\pdfoutput=1
\usepackage[utf8x]{inputenc}
\usepackage[T1]{fontenc}

\usepackage[modulo,switch]{lineno}
\modulolinenumbers[1]

\usepackage{amssymb,amsmath,mathtools,amsfonts, cuted}
\usepackage{graphicx}
\usepackage{float}
\usepackage[colorinlistoftodos]{todonotes}
\usepackage[colorlinks=true, allcolors=cyan]{hyperref}
\usepackage{subcaption}
\usepackage{tabularx,booktabs}
\usepackage{xcolor}

\graphicspath{{./img/}}

\makeatletter\@ifundefined{date}{}{\date{}}
\makeatother

\evensidemargin\oddsidemargin \hoffset-22.4mm \voffset-28.4mm
\begin{document}

\thispagestyle{empty}
\title{Amplitude-dependent modal coefficients accounting for localized nonlinear losses in a time-domain integration of woodwind model}

\author{Nathan Szwarcberg \textsuperscript{a, b}, Tom Colinot \textsuperscript{a, b}, Christophe Vergez \textsuperscript{b}, Michaël Jousserand \textsuperscript{a} \\
\textsuperscript{a} Buffet Crampon, 5 Rue Maurice Berteaux, 78711 Mantes-la-Ville, France\\
\textsuperscript{b} Aix Marseille Univ, CNRS, Centrale Marseille, LMA, Marseille, France
}

\maketitle\thispagestyle{empty}

\begin{abstract}
This article develops the design of a sound synthesis model of a woodwind instrument by modal decomposition of the input impedance, taking into account viscothermal losses as well as localized nonlinear losses at the end of the resonator.
This formalism has already been applied by Diab et al. (2022) to the study of forced systems. It is now implemented for self-oscillating systems.
The employed method extends the definition of the input impedance to the nonlinear domain by adding a dependance on the RMS acoustic velocity at a geometric discontinuity. The poles and residues resulting from the modal decomposition are fitted as a function of this velocity. Thus, the pressure-flow relation defined by the resonator is completed by new equations which account for the dependence with the velocity at the end of the tube.
To assess the ability of the model to reproduce a real phenomenon, comparisons with the experimental results of Atig et al. (2004) and Dalmont et al. (2007) were carried out.
Simulations show that the model reproduces these experimental results qualitatively and quantitatively.
\end{abstract}

\section{Introduction}

Sound synthesis by modal decomposition is based on an input impedance measurement, which captures the passive acoustical response of a real instrument.
This method thus has the advantage of capturing the acoustical subtleties that differentiate two clarinets by means of a single measurement. 
Other methods, such as delay lines \cite{guillemain2005real}, waveguides \cite{scavone1997acoustic,mignot2009phd} or spatial discretization \cite{bilbao2009numerical, ernoult2021full} would require more effort on the geometrical description of the resonator to render such acoustical subtleties.

{Furthermore, sound synthesis by modal decomposition requires little RAM compared to such other methods, which can be beneficial for embedded devices.
Indeed, in modal decomposition synthesis, the temporal integration scheme requires to keep only a few previous iterations (precisely two for \cite{guillemain2005real}) to compute a new one. For the delay line synthesis, it is necessary to keep in memory all the temporal iterations during a round trip, i.e. during $2 L F_s / c_0$ iterations for an academic cylindrical resonator  without lateral holes. 
Although this is not a problem for most embedded processors, fine modeling of viscothermal losses in the waveguide formalism also requires additional computations involving fractional derivatives \cite{bilbao2013finite}.
Finally, the ODE formalism is particularly well suited to the bifurcation analysis, which is not the case of the PDE \cite{bilbao2013finite} or of the DDE  formalisms \cite{guillemain2006digital}}.

The input impedance is a measure of the linear frequency response of the resonator, for a low-amplitude excitation. However, in the case of woodwind instrument playing, the measured acoustic pressure and velocity in the resonator can be very high. For instance, \cite{bergeot2014response} measure a pressure in the mouthpiece of 4~kPa, i.e. 163~dB for a monochromatic wave. According to \cite[Chap. 8.4.5]{bible2016}, from an acoustic speed of roughly 1 m/s, jet separation phenomena appear. At the level of a geometrical discontinuity, such as the open end of a pipe \cite{disselhorst1980flow} or a lateral orifice \cite{ingaard1950acoustic}, vortices are observed. A part of the kinetic energy of the jet is absorbed by these vortices and is dissipated as heat by friction. These losses are accounted for through a nonlinear relationship derived from Bernoulli's law, as demonstrated by the implementation in waveguide modeling or delay lines of \cite{atig2004saturation}, \cite{guillemain2006digital}, \cite{dalmont_oscillation_2007} and \cite{taillard2018phd}.

Following \cite{ingaard1950acoustic}, the works from Atig \cite{atig2004saturation, atig2004termination} and Dalmont \cite{dalmont_experimental_2002, dalmont_oscillation_2007} model localized nonlinear losses through a resistive impedance $Z_t$. 
This impedance is in series with the radiation impedance $Z_R$, and depends on the amplitude of the acoustic velocity {$v_{0}$} { at the location of the discontinuity:}
\begin{equation}\label{eq_1}
Z_t = \frac{4 c_d}{3 \pi} \frac{| {v_{0}}|}{c_0} Z_c,
\end{equation}
where $Z_c=\rho_0 c_0/S$ is the characteristic impedance of the medium for plane waves, and $c_d \in [0;3]$ is a parameter depending on the geometry of the termination. 
The larger the radius of curvature at the output, the smaller this coefficient is. 
The coefficient $c_d$ is difficult to predict theoretically, and easier to determine experimentally \cite{atig2004saturation}. As shown by \cite{atig2004saturation}, nonlinear losses have a significant influence on the playing range of a clarinet, hence models of sound production in woodwinds should include such effects.

Models of woodwind instruments taking into account localized nonlinear losses at the end of the resonator have never been developped in the framework of modal synthesis. This is the main contribution of this work.

First, a method proposed by \cite{diab2022nonlinear} to account for nonlinear losses by modal decomposition is presented. It is then applied to a cylindrical tube with nonlinear losses located at the open end. From this resonator, a self-oscillating clarinet-like system is defined and simulated by time integration. These simulations are finally compared to the experimental results published by \cite{atig2004saturation} and \cite{dalmont_oscillation_2007}.

\section{Method of modal decomposition accounting for localized nonlinear losses}
The definition of a "nonlinear impedance" (i.e. adjusting the impedance so that the link between acoustic flow and pressure stays valid in nonlinear conditions) opens the way to consider localized nonlinear losses by modal decomposition. 
In \cite{diab2022nonlinear}, the evolution of the surface impedance of perforated plates is computed by temporal simulation for a broadband excitation, with respect to the RMS amplitude of the acoustic velocity at the level of the hole, noted $v_{RMS}$. 
From the decomposition of the impedance in the linear domain into a sum of $N$ modes, two methods are studied for increasing values of $v_{RMS}$: the interpolation of the impedance, on one hand, and the regression of the poles and complex residues $(s_n,C_n)$ on the other hand. 
This second method will be considered hereafter. 

Nonlinear effects are taken into account by allowing the poles and residues to vary with respect to $v_{RMS}$. In the Laplace domain, the modal decomposition of the input impedance {is written}:
\begin{multline}\label{eq_zin_v}
Z(s,v_{RMS}) =Z_c \sum_{n=1}^N \frac{C_n(v_{RMS})}{s - s_n(v_{RMS})} \\
   + \frac{C_n^*(v_{RMS})}{s - s_n^*(v_{RMS})},
\end{multline}
where $s$ is the Laplace variable and $\bullet^*$ denotes the complex conjugation operation. The relationship between poles and residues and $v_{RMS}$ is assumed to be a rational fraction \cite{diab2022nonlinear}. In the present work, the approximation is limited to a polynomial of degree $N_p$:
\begin{align}\label{eq_ex_interp}
C_n(v_{RMS}) &= C_n^{(0)} + \sum_{k=1}^{N_p}  C_n^{(k)} \left(v_{RMS} \right)^k, \\
s_n(v_{RMS}) &= s_n^{(0)} + \sum_{k=1}^{N_p}  s_n^{(k)} \left(v_{RMS} \right)^k, 
\end{align}
where the index $\bullet^{(0)}$ refers to the linear part of the modal coefficient. In \cite{diab2022nonlinear}, the regression coefficients of the poles and residues are then adjusted on a nonlinear surface impedance model $Z_{NL}(v_{RMS})$, given by \cite{laly2018acoustical}. It is worth noting that in the present paper, the poles and residues are fitted on the dimensioned quantity of the RMS velocity. Therefore, the unit of coefficients $C_n^{(k)}$ and $s_n^{(k)}$ are expressed in $\mathrm{rad} \cdot \mathrm{s}^{-1} \cdot [\mathrm{m} \cdot \mathrm{s}^{-1}]^{-k}$.

The last step consists in computing $v_{RMS}$, given by:
\begin{equation}
v_{RMS}^2(t) = \frac{1}{t} \int_0^t {v_{0}}^2(\tau)\mathrm{d}\tau,
\end{equation}
where ${v_{0}}(t)={u_{0}}/S$ is the acoustic velocity at the {discontinuity}, of cross section $S$. This definition is rather impractical in the framework of a time-domain simulation. In order to obtain a state formalism, it is replaced by:
\begin{equation}\label{eq_vrms}
\frac{\partial (t v^2_{RMS})}{\partial t} = {v_{0}}(t)^2,
\end{equation}
which is equivalent. 
By considering the right-hand side of Eq. \eqref{eq_vrms} as an input for the ODE solver, $tv_{RMS}^2$ can be computed, hence $v_{RMS}$.

\section{Application to a cylindrical tube with nonlinear losses at the open end}\label{sec_2}

The impedance regression technique presented by \cite{diab2022nonlinear} is applied to the case of a closed-open cylinder of radius $R$ and length $L$, which is a similar case study to \cite{atig2004saturation}. Viscothermal losses are taken into account in the propagation: 
\begin{equation}\label{eq_Gamma}
\Gamma(s)=\frac{s}{c_0} + (1+j)\frac{\eta \sqrt{s}}{R \sqrt{2 j \pi}},
\end{equation}
where $\eta=3 \cdot 10^{-5}~\mathrm{s}^{1/2}$.
The boundary condition at $x=L$ is given by the radiation impedance $Z_R$, which includes in series the nonlinear impedance $Z_t$ defined by Eq. \eqref{eq_1}: 
\begin{equation}\label{eq_zr}
\begin{split}
Z_R &=   Z_R^{(lin)} + Z_t, \\
\mathrm{where} \quad Z_R^{(lin)} &=Z_c \left( jk\Delta l + \frac{1}{4}(kR)^2 \right), \\
Z_t &= Z_c  \frac{v_{RMS}(L,t)}{c_0}K_{NL},
\end{split}
\end{equation}
and $jk=s/c_0$, $\Delta l \approx 0.6R$ \cite[Chap. 12.6.1.3]{bible2016},  $K_{NL}=4 c_d/ (3 \pi)$. In the following, $v_{RMS}(L,t)$ will be denoted as $v_{RMS}$ for reading comfort. 
The input impedance is defined by:
\begin{equation}\label{eq_Zin}
z_{in} = \tanh \left( \Gamma L + \tanh^{-1}(z_R ) \right),
\end{equation}
introducing the dimensionless notation $z_\bullet~=~Z_\bullet/Z_c$. A linear input impedance $z_{in}^{(lin)}$ is also defined, such that 
\begin{multline}\label{eq_zin_lin}
z_{in}^{(lin)} = \tanh \bigl( \Gamma L \\
+ \tanh^{-1}\bigl[ jk\Delta l + \frac{1}{4}(kR)^2 \bigr] \bigr),
\end{multline}
which is equivalent to $z_{in}$ for $v_{RMS}=0$.

\begin{figure*}[h!]
	\centering
	\includegraphics[width=.9\textwidth]{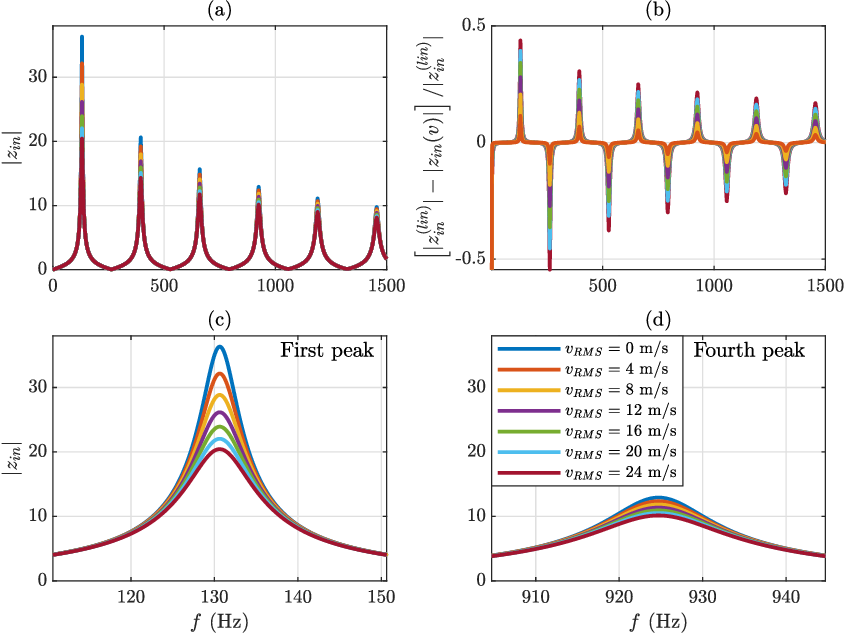}
	\caption{Input impedance of a cylinder $z_{in}$ (computed through Eq. \eqref{eq_Zin} and Eq. \eqref{eq_zin_lin}) of length $L=64$~cm and radius $R=8$~mm for different values of the acoustic RMS velocity at the open end. {The value of $c_d$ has been set to $13/9$, which corresponds to a radius of curvature of 0.3~mm, according to \cite{atig2004phd}}. From top to bottom, from left to right : modulus of the input impedance; relative gap between nonlinear and linear definition of $z_{in}$; detail view on the first peak; detail view on the fourth peak.}
	\label{fig_1}
\end{figure*}

Figure \ref{fig_1} illustrates the evolution of $z_{in}$ (computed through Eq. \eqref{eq_Zin} and Eq. \eqref{eq_zin_lin}, in which the Laplace variable is substituted by $j2 \pi f$) as a function of the RMS velocity at the open end. 
It can be observed that taking into account nonlinear losses has a consequence on the amplitude, but not on the frequency of the resonance peaks.
{ Figure \ref{fig_1}~b} shows that this impact is particularly accentuated on the first resonance peak: its amplitude is 43~\% lower for $v_{RMS}=24$~m/s than in the case without nonlinear losses. 
The difference is reduced to 30~\% for the second peak, and to 16~\% for the sixth peak.
Near the antiresonances, the negative relative discrepancies are high, since the impedance values are all close to zero. The absolute deviation remains very low. From the lack of frequency variation of the resonance peaks with $v_{RMS}$, we could anticipate that nonlinear losses at the end of the pipe would rather impact dynamics than intonation.

\subsection{Modal decomposition of the nonlinear input impedance}

In order to take into account nonlinear losses in temporal simulation, the modal coefficients of the input impedance must be determined. In the same way that the nonlinear input impedance depends on the acoustic velocity at the open end, the coefficients $C_n$ and $s_n$ derived from the modal decomposition of $z_{in}$ also depend on $v_{RMS}$. 
{To formulate the relationship between the modal coefficients and $v_{RMS}$, a minor adaptation of the method presented in \cite[Chap. 5.5.3]{bible2016} is used. The definition of $z_R$ used in \cite{bible2016} is substituted by the expression given by Eq. \eqref{eq_zr}, taking into account nonlinear losses at the end of the tube.}

\subsubsection{Expression of the poles $s_n$}
 
\begin{figure*}[h!]
	\centering	
	\begin{subfigure}[t]{0.45\textwidth}
		\centering
		\includegraphics[width = \textwidth]{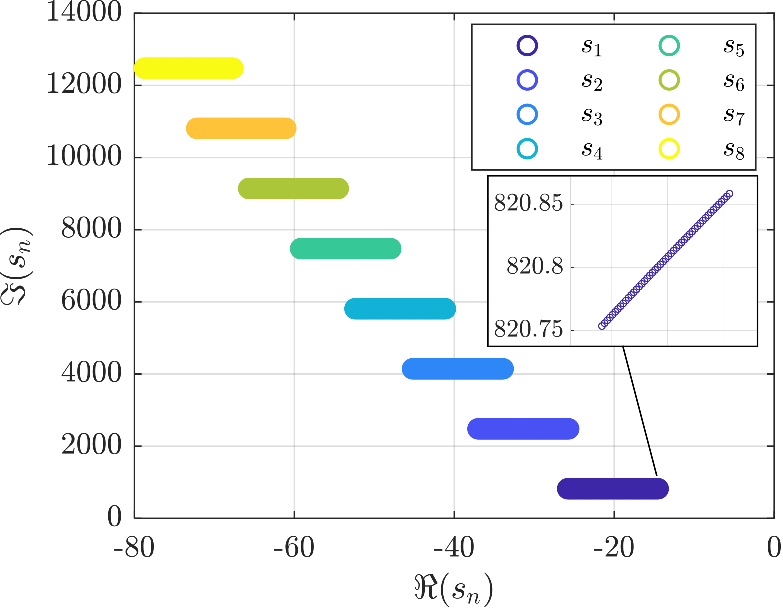}
		\caption{Real and imaginary parts of poles $s_n(v_{RMS})$, calculated by solving Eq. \eqref{eq_sn}. Detailed view on $s_1$.}
		\label{fig_2_1}
	\end{subfigure}
	\hfill
	\begin{subfigure}[t]{0.45\textwidth}
	\centering
		\includegraphics[width = \textwidth]{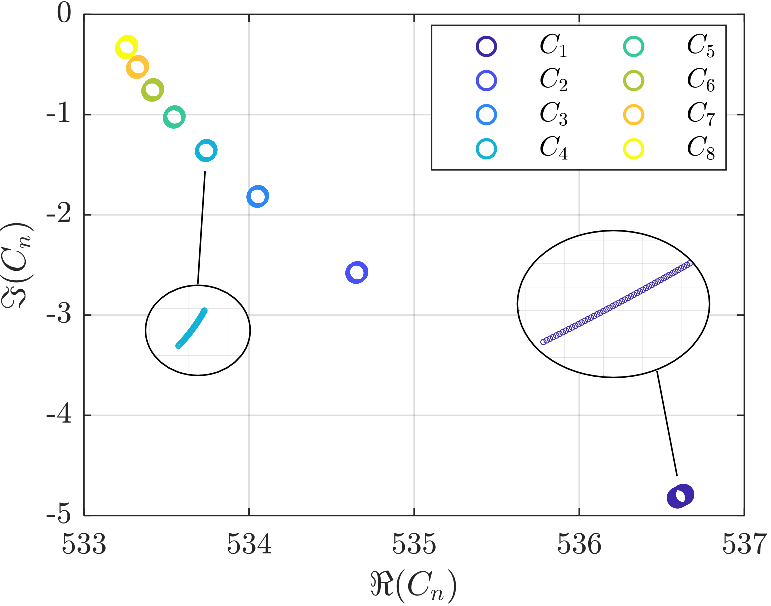}
		\caption{Real and imaginary parts of residues $C_n(v_{RMS})$, calculated with Eq. \eqref{eq_Cn}. Detailed views on $C_1$ and $C_4$.}
		\label{fig_2_2}
	\end{subfigure}
	\caption{Graphical representation of the $N=8$ first poles and residues. The values of $v_{RMS}$ are linearly chosen between 0~m/s and 24~m/s, according to \cite{atig2004phd}. {As in Figure \ref{fig_1}, $c_d=13/9$.}}
	\label{fig_2}
\end{figure*}

Eq. \eqref{eq_Zin} can also be written as:
\begin{equation}
\begin{split} 
z_{in} =\frac{\sinh \left[ \Gamma(s) L + h(s,v_{RMS}) \right]}{\cosh \left[ \Gamma(s) L + h(s,v_{RMS}) \right]}, \\
\mathrm{where} \quad h(s,v_{RMS})=\tanh^{-1}(z_R).
\end{split}
\end{equation}
The poles $s_n(v_{RMS})$ are the solutions of $\cosh \left[ \Gamma L + h\right]=0$, i.e.:
\begin{equation}\label{eq_sn}
\Gamma(s_n) L + h(s_n,v_{RMS})-j \frac{(2n-1)\pi}{2}=0 .
\end{equation}
The solutions of Eq. \eqref{eq_sn} give $s_n$, for different input values of $v_{RMS}$. They are plotted on Figure \ref{fig_2_1}. {The consideration of} nonlinear losses at the end of a cylindrical pipe has almost no influence on $\Im(s_n)$, which is the resonance {angular frequency} of peak $n$. { However, a larger $v_{RMS}$ increases $|\Re (s_n)|$}.
This shift remains almost the same regardless of the index of the pole. However, in terms of relative deviation, the ratio $|\Re(s_n-s_n^{(lin)}) / \Im(s_n) |$ is decreasing as $n$ increases.
This is signaled by the amplitude of peaks of higher index appearing less affected by nonlinear losses, as shown in {Figure \ref{fig_1}~b}. 

\subsubsection{Expression of the residues $C_n$}

It remains to calculate residue $C_n(v_{RMS})$. In the vicinity of a pole $s_n$, the denominator $D$ of $z_{in}$ can be written through a first-order series expansion of $\cosh$: 
\begin{equation}
\begin{split}
	D &= (s-s_n)D'(s_n,v_{RMS}) \\
	\mathrm{where} \quad  D' &= (\Gamma' L + h' ) \sinh\left( \Gamma L + h \right),
\end{split}
\end{equation}
introducing notation $\bullet'=\partial \bullet / \partial s $. According to \cite[Chap. 5.5.3]{bible2016}, after application of the residues theorem, the modal decomposition of the input impedance for the $n$-th peak can be written as
\begin{equation}\label{eq_Cn}
\begin{split}
Z_{in, n} &=\frac{P_n(s)}{U(s)} = Z_c\frac{C_n}{s-s_n},~ \mathrm{where} \\
  C_n(v_{RMS}) &= \frac{1}{\Gamma'(s_n)L + h'(s_n,v_{RMS})}.
\end{split}
\end{equation}
The evolution of coefficients $C_n$ with respect to $v_{RMS}$ is plotted on Figure \ref{fig_2_2}. Real and imaginary parts of $C_n$ slightly decrease when $v_{RMS}$ increases.
Following \cite{diab2022nonlinear}, it remains to fit the modal coefficients with respect to $v_{RMS}$, as in the example of Eq.  \eqref{eq_ex_interp}. 

\subsubsection{Polynomial fitting of the modal coefficients}
\begin{figure*}[h!]
	\centering
	\begin{subfigure}[t]{.45 \textwidth}
		\centering
		\includegraphics[width = \textwidth]{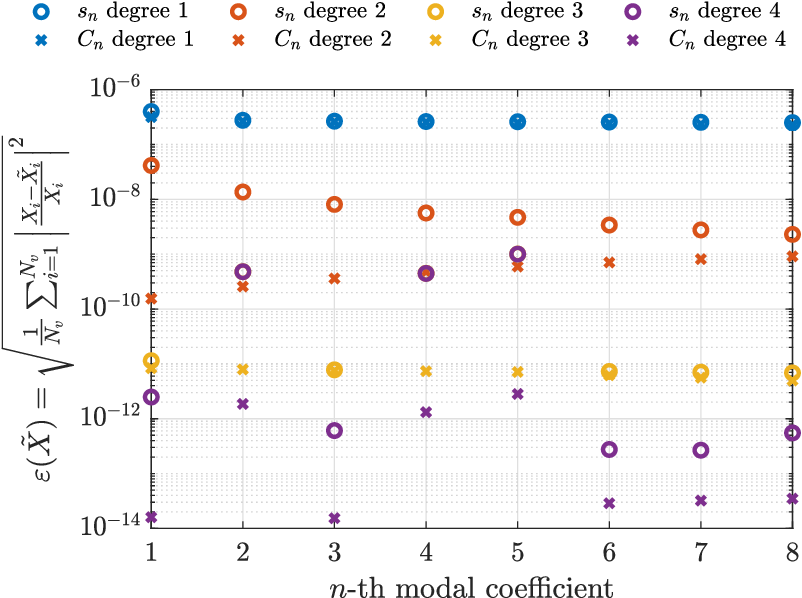}
		\caption{Mean relative error $\varepsilon$ between fitted modal coefficient $\tilde{X}= \{\tilde{s}_n, \tilde{C}_n\}$ and its actual value $X=\{s_n,C_n\}$. Regression is based on $N_v=25$ values of $v_{RMS}$.}
		\label{fig_3_1}
	\end{subfigure}
	\hfill
	\begin{subfigure}[t]{.45 \textwidth}
		\centering
		\includegraphics[width = \textwidth]{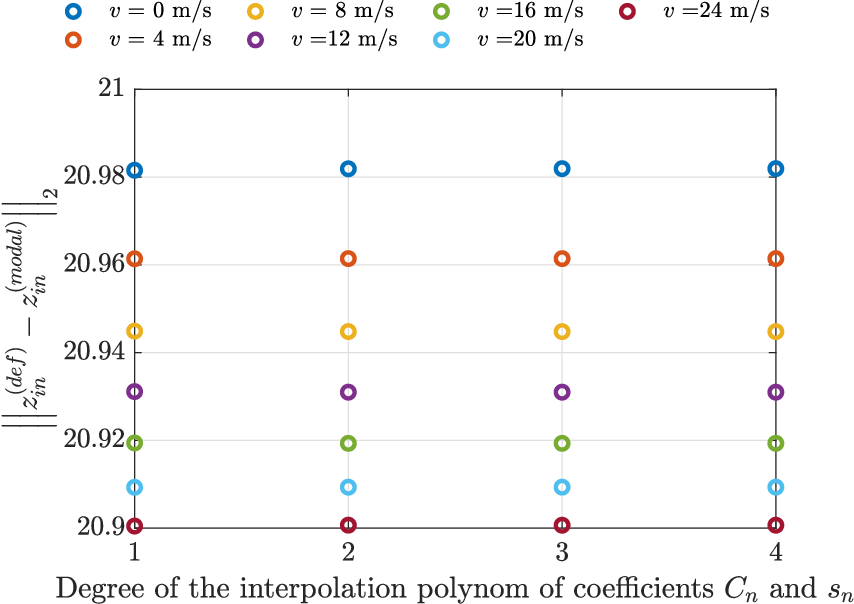}
		\caption{$L_2$-norm of the error over $z_{in}$ for every frequencies, between definition $z_{in}^{(def)}$ given by Eq. \eqref{eq_Zin} and modal decomposition approximation $z_{in}^{(modal)}$ defined by Eq. \eqref{eq_zin_v}. This error is plotted with respect to the degree of the fitting {polynomial} of $C_n$ and $s_n$.}
		\label{fig_3_2}
	\end{subfigure}
	\caption{Error due to the regression of modal coefficients $C_n$ and $s_n$ (Figure \ref{fig_3_1}), and its consequence on the discrepancy between modal decomposition and definition \eqref{eq_Zin} on $z_{in}$ (Figure \ref{fig_3_2}).}
	\label{fig_3}
\end{figure*}

Figure \ref{fig_3_1} shows the mean relative fitting error for {polynomials} of different degrees, for the same data as in Figure \ref{fig_2}. 
For linear regression, the mean relative error stays around $10^{-7}$ for each mode $n$, both for $C_n$ and $s_n$. Further simulations reveal that choosing an excessive values of $c_d$ tends to increase this error. 
For a tube with sharp edges at the open end, \cite{atig2004saturation} estimated a maximum value of $c_d=2.8$.
For $c_d=5$, for instance, the mean relative error of linear regression is around $10^{-5}$. Although this error has increased, it remains very low.

Moreover, Figure \ref{fig_3_2} shows that choosing an polynomial regression degree larger than 1 has no consequence on the error over $z_{in}$ between modal decomposition with fitted coefficients and {the} definition given by Eq. \eqref{eq_Zin}.
This error is therefore inherent to the modal decomposition approximation (Eq. \eqref{eq_zin_v}) of the input impedance (Eq. \eqref{eq_Zin}).

According to these previous results, in the rest of this article, the modal coefficients will both be fitted by a polynomial of order 1.

\section{Application to the time-integration simulation of a clarinet-like system}

The cylindrical tube considered in section \ref{sec_2} is now treated as the resonator of a clarinet-like instrument.
The physical model governing the self-sustained oscillations of the instrument is presented in section \ref{section_physical model}. 
It is then simulated by time integration, imposing a linear increase of the blowing pressure (\textit{crescendo}) followed by a linear decrease (\textit{diminuendo}). Details and results from the simulations are given in section \ref{section_simulation}.

\subsection{Physical model for the self-sustained oscillations}\label{section_physical model}
The self-sustained system consists of three main parts: reed dynamics, reed channel and resonator. 
This last subsystem is enriched compared to the classical modal decomposition formalism \cite{dubos1999theory,silva2007simulation,taillard2018modal} by the addition of nonlinear losses at the end of the pipe.

\subsubsection{Reed dynamics}\label{section_reed_dyn}
The dynamic behavior of the reed is modeled for its dimensionless displacement $x(t)$ by the following equation :
\begin{multline}
\dfrac{1}{\omega_r^2}\ddot{x}(t) +\dfrac{q_r}{\omega_r}\dot{x}(t)+ x(t)=p(t) \\
- \gamma(t) + F_c(x, \dot{x}),
\end{multline}
where $\omega_r=2\pi \times 2200$~rad/s  is the reed resonance {angular frequency} \cite{chatziioannou2012estimation} and $q_r=0.4$  is the reed damping \cite{silva_approximation_2009}. 
$\gamma(t)=P_m(t)/p_M$ is the dimensionless blowing parameter and $p(t)=p^{(dim)}(t)/p_M$ is the dimensionless pressure at the input of the resonator. 
The beating pressure $p_M$ is the value of the blowing pressure $P_m$ for which the reed closes the reed channel, in quasistatic regime.
$F_c(x, \dot{x})$ is the contact force function. In the following, a model of "ghost reed" \cite{colinot2019ghost} is considered, i.e. $F_c=0$.

\subsubsection{Reed channel}\label{section_reedchannel}
The characteristic of the flow $u(t)$ {through} the reed channel is described by the following equation, according to \cite{wilson_operating_1974}:
\begin{multline}\label{eq_u}
u(t)=-\lambda \dot{x}(t) +  \zeta \left[x(t)+1\right]^+ ~\mathrm{sgn} \bigl[\gamma(t) \\
-p(t)\bigr] \sqrt{\left|\gamma(t)-p(t)\right|},
\end{multline}
where $\lambda=5.5\cdot 10^{-3}/c_0$ is the reed flow parameter \cite{dalmont2003nonlinear}, $\zeta$ is the embouchure parameter, and $\gamma(t)$ is the dimensionless blowing pressure. The operator $[\bullet]^+=(\bullet + |\bullet|)/2$ refers to the positive part function. For the simulations, the absolute value functions are regularized to smooth the irregularities of the quasi-static flow characteristic, without altering its behavior, according to \cite{colinot_numerical_2020}. Thus, $|\bullet|\mapsto \sqrt{\bullet^2 + \eta}$, with $\eta=0.001$.
\subsubsection{Resonator}\label{section_resonator}
The resonator is described in the frequency domain by its input impedance $z_{in}$, under its modal decomposition form given by Eq. \eqref{eq_zin_v}. 
The modal poles and residues $s_n$ and $C_n$ are chosen to be linearly dependent on $v_{RMS}$, i.e.:
\begin{align*}
C_n(v_{RMS}) &= C_n^{(0)} + C_n^{(1)} v_{RMS}, \\
s_n(v_{RMS}) &= s_n^{(0)} + s_n^{(1)} v_{RMS}.
\end{align*}
The modal decomposition of $z_{in}$ allows to write the relation between the dimensionless pressure $p$ and flow $u$ in the temporal domain:
\begin{multline}\label{eq_zin_t}
\dot{p}_n(t) - s_n(v_{RMS}) p_n(t) = C_n(v_{RMS}) u(t), \\
\mathrm{where} \quad  p(t) = 2 \sum_n^N \Re(p_n(t)).
\end{multline}
The values of $s_n$ and $C_n$ must be updated at each time step by computing $v_{RMS}(L,t)$. The method for calculating the mean velocity at nonlinearity is detailed in the following section.

\subsubsection{Computation of $v_{RMS}$ at each time step}\label{section_vrms}

To compute $v_{RMS}(L,t)$, the pressure at the termination $L$ must be calculated. 
First, a linear problem is considered. 
The value of $s_n$ is therefore chosen for $v_{RMS}=0$~m/s. The modal components of the pressure are related to the total pressure at $L$ through the following equation:
\begin{equation}
\begin{split}
&p(L,t) = 2\Re \left( \sum_n^N p_n(t) \phi_n(L) \right), \\
&\mathrm{where} \quad \phi_n(\xi) = \cosh(\Gamma(s_n^{(0)}) \xi),
\end{split}
\end{equation}
and $\xi\in[0,L]$ is the distance along the longitudinal axis of the resonator.
For reading comfort, the expression $p(t)$ will be {reserved} to denote $p(0,t)$.

The dimensionless acoustic velocity at the open end is then calculated from its definition which is related to the pressure field through the dimensionless Euler equation:
\begin{equation}
\frac{\partial p}{\partial \xi}(\xi,t) = - \frac{1}{c_0} \frac{\partial v}{\partial t}(\xi,t).
\end{equation}
The dimensionless acoustic velocity $v(L,t)$ is thus obtained by numerical integration of the following equation :
\begin{multline}\label{eq_v}
 \dot{v}(L,t) = - 2 c_0 \Re  \sum_n^N p_n(t) \frac{\mathrm{d} \phi_n}{\mathrm{d} \xi}(L), \\
 \mathrm{where} \quad  \frac{\mathrm{d} \phi_n}{\mathrm{d} \xi}(\xi) = \Gamma(s_n^{(0)}) \sinh\left[\Gamma(s_n^{(0)})\xi \right] .
\end{multline}
The RMS velocity is finally obtained by double time integration of Eq. \eqref{eq_v}, according to Eq. \eqref{eq_vrms}. 
Since the equations of the system characterizing the self-sustained oscillations are dimensionless, it is necessary to resize $v_{RMS}$, because the modal coefficients $C_n^{(k)}$ and $s_n^{(k)}$ presented in Eq. \eqref{eq_ex_interp} have been regressed from the dimensioned velocity.
 To do so, $v_{RMS}^{({adim})}$ is multiplied by $p_M/(\rho_0 c_0)$. 
 In the following simulations, a value of $p_M=8.5$~kPa is set, according to experimental results from \cite{atig2004saturation}.

Regarding the calculation of $v(L,t)$, it is not numerically guaranteed that the mean value of $\dot{v}(L,t)$ remains zero. Consequently, the integral $v(L,t) = \int_0^t \dot{v}(L,t) \mathrm{d}t$ may diverge. To avoid these divergence problems when integrating $\dot{v}(L,t)$ and $v^2(L,t)$, a short-memory term $\tau$ is added within each integral. Thus, $v(L,t)$ is now defined by the following convolution product:
\begin{equation}\label{eq_v_memory}
v(L,t) = \int_0^t \dot{v}(L,u) e^{-\frac{t-u}{\tau}} ~\mathrm{d} u,
\end{equation}
which {is written}, in the Laplace ($\mathcal{L}$) domain, as:
\begin{equation}
\mathcal{L}\left[ v(L,t) \right] = \mathcal{L}\left[ \dot{v}(L,t) \right] \dfrac{1}{s+1/\tau}.
\end{equation}
By going back to the time domain, Eq. \eqref{eq_v_memory} becomes :
\begin{multline}
\dot{v}(L,t) = - 2 c_0 \Re \left( \sum_n^N p_n(t)  \frac{\mathrm{d}\phi_n}{\mathrm{d} \xi}(L) \right) \\
- \frac{1}{\tau}v(L,t).
\end{multline}
The same method is applied for the computation of $v_{RMS}$:
\begin{align}
v^2_{RMS} &= \frac{1}{\tau} \int_0^t e^{-\frac{t-u}{\tau}}v^2(L,u)~\mathrm{d} u \\
\Leftrightarrow \mathcal{L}\left[ v^2_{RMS} \right] &= \frac{1}{\tau} \mathcal{L}\left[ {v}^2(L,t) \right] \frac{1}{s+1/\tau} \\
\Leftrightarrow \frac{\partial(\tau v^2_{RMS})}{\partial t} &= v^2(L,t) - v^2_{RMS}.
\end{align}

\subsubsection{Complete system of equations}
The complete self-sustained system is governed by the equations defined in sections \ref{section_reed_dyn}, \ref{section_reedchannel}, \ref{section_resonator} and \ref{section_vrms}. 
This article shows the resolution of this problem using an ode solver. 
The Cauchy problem $\dot{\mathbf{Y}}=\mathcal{F}(\mathbf{Y},t)$ writes, according to Eq. \eqref{eq_system}, as: 
\begin{equation}
\mathcal{F}(\mathbf{Y},t) = \mathcal{F}\left(\begin{matrix}
x(t) \\ \dot{x}(t) \\ v(L,t) \\ \tau v^2_{RMS} \\p_1(t) \\ \vdots \\p_N(t)
\end{matrix} \right)
\end{equation}
\begin{equation}\label{eq_system}
=\left(	\begin{matrix}
\dot{x}(t) \\ 

-q_r \omega_r \dot{x}(t)  + \omega_r^2 \left[ p(t) - \gamma(t) - x(t)\right] \\

- 2 c_0 \Re \left( \displaystyle \sum_n^N p_n(t)  \frac{\mathrm{d}\phi_n}{\mathrm{d} \xi}(L) \right) - \frac{1}{\tau}v(L,t) \\

v^2(L,t) - v^2_{RMS} \\

C_1(v_{RMS})u(t) + s_1(v_{RMS})p_1(t) \\

\vdots \\

C_N(v_{RMS})u(t) + s_N(v_{RMS})p_N(t)	
\end{matrix} \right),
\end{equation}
where $u(t)$ is computed by using Eq. \eqref{eq_u}, $C_k(v_{RMS})$ are computed with Eq. \eqref{eq_Cn}, and $s_k(v_{RMS})$ are computed with Eq. \eqref{eq_sn}. The initial conditions $\mathbf{Y_0}$ will be set to $\mathbf{0}$ in the following simulations.

\subsection{Simulation including nonlinear losses at the open end}\label{section_simulation}

The model including nonlinear losses at the open end as represented by the system of Eq. \eqref{eq_system} is now simulated by time integration, using the solver \texttt{ode45} from \textsc{Matlab}. The input data of the problem are based on experimental results from \cite{atig2004saturation}.

\subsubsection{Simulation parameters}\label{section_sim_param}
The parameters related to the resonator, the reed and the reed channel are presented in table \ref{tab_1}. Six values of $c_d$ have been taken from \cite{atig2004saturation}, which are between $c_d=0$ and $c_d=2.8$. 
The value of $\zeta=0.28$ was calculated using data corresponding to a "loose embouchure" configuration.
For the calculation of $v_{RMS}$, the short-memory term was set to the period of the first impedance peak, i.e. $\tau = 2 \pi / \Im(s_1)$. 
The evolution of $\gamma$ is first linearly ascending, from $\gamma=0$ to $\gamma=3$ in 8~s, then linearly descending, from $\gamma=3$ to $\gamma=0$ in 8~s. 
The total simulation time is therefore 16~s.
{Simulations were performed for $N=4$ modes. A convergence study showed that the dynamic behavior of the system did not change for a higher number of modes.}

\begin{table}[H]
	\centering
	\caption{Main constants used for the simulations.}
	\label{tab_1}
	\begin{tabular}{cccc}
	\toprule
	$L$ (cm) & $R$ (mm) & $N$ modes &$\omega_r/2\pi$ (Hz)  \\
	64 & 8 & 4 & 2200 \\ 
	\midrule
	$q_r$ & $\lambda$ (s) & $\zeta$ & $p_M$ (kPa) \\
	0.4 & $1.6\cdot 10^{-5}$ & 0.28 & 8.5\\
	\bottomrule
	\end{tabular}
\end{table}

\subsubsection{Results}
\begin{figure}[H]
	\centering
	\begin{subfigure}[c]{0.5 \textwidth}
		\centering
		\includegraphics[width=.95\textwidth]{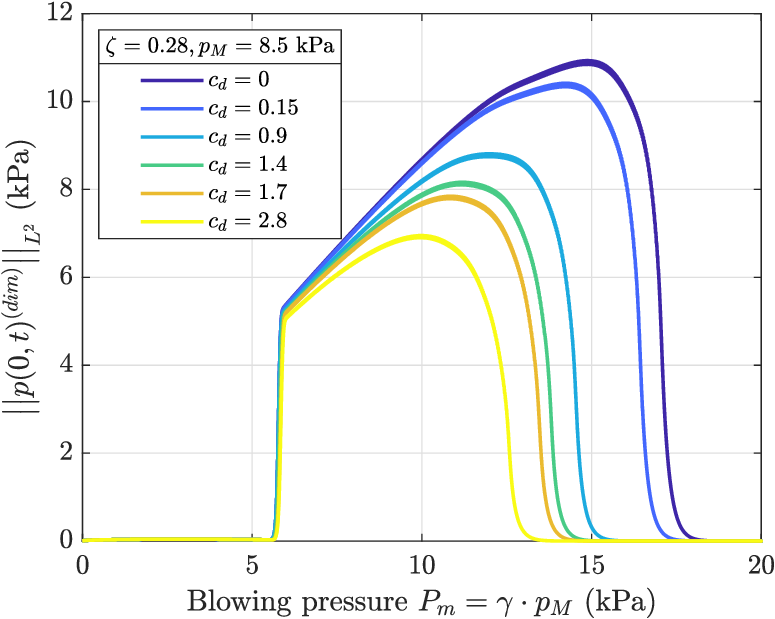}
		\caption{\textit{Crescendo}.}
		\label{fig_4_1}
	\end{subfigure}
	\hfill	
	\begin{subfigure}[c]{0.5 \textwidth}
		\centering
		\includegraphics[width=.95\textwidth]{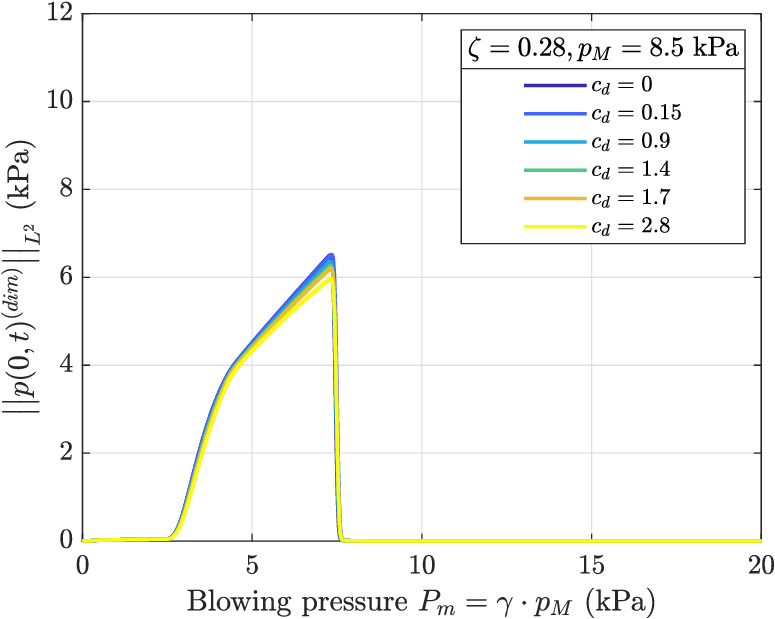}
		\caption{\textit{Diminuendo}.}
		\label{fig_4_2}
	\end{subfigure}
	\caption{Bifurcation diagram of the L$_2$ norm of $p^{(dim)}$ with respect to $P_m$, for different values of $c_d$, in an ascending blowing pressure configuration (\ref{fig_4_1}) and a descending one (\ref{fig_4_2}).}
	\label{fig_4}
\end{figure}

The bifurcation diagram of the input pressure $p^{(dim)}=p \cdot p_M$ over the blowing pressure $P_m=\gamma \cdot p_M$ is represented on Figure~\ref{fig_4}, with the parameters detailed in section~\ref{section_sim_param}.

\begin{table}[H]
\centering	
\caption{Values of the extinction threshold { $P_{Íextup}=p_M \cdot \gamma_{extup}$ (kPa)}, for the simulation of Figure \ref{fig_4_1}.}
\label{tab_2}
\begin{tabular}{lcccccc}
\toprule
$c_d$ & 0 & 0.15 & 0.9 & 1.4 & 1.7 & 2.8 \\
 $ P_{extup}$ & 17.7 & 17.3 & 15.3 & 14.5 & 14.2 & 13.2 \\
\bottomrule
\end{tabular}
\end{table}

During a \textit{crescendo} (Figure \ref{fig_4_1}), the oscillation threshold is located near 5.7~kPa ($\gamma_{th} \approx 0.67$).
This threshold remains the same, whether  nonlinear losses are taken into account ($c_d\neq 0$) or not ($c_d=0$). 
However, nonlinear losses have a significant influence on the extinction threshold $\gamma_{extup}$, i.e. the blowing pressure from which the reed stops oscillating and is pressed completely against the mouthpiece. This influence is shown in Table~\ref{tab_2}. 
When $c_d$ is increased (i.e. nonlinear losses increase), $\gamma_{extup}$ diminishes. 
Similarly to the experiments conducted by \cite{atig2004saturation}, nonlinear losses have an important influence on the dynamic playing range of the musician.
{Figure \ref{fig_4} shows that when nonlinear losses are low, the range of stable oscillation amplitude that can be obtained is larger.}

In the \textit{diminuendo} phase, the inverse threshold is the same ($\gamma_{inv}\approx 0.89$) for each geometry at the open end of the pipe. 
Around this threshold, the amplitude of the input pressure decreases slightly as the losses increase. 
This behavior is also observed in the experimental curves of \cite[Figure 12 a, c and e]{atig2004saturation}. 

This comparaison with the experimental results of Atig et al. is limited to a qualitative study. 
Parameters such as the reed resonance {angular frequency} $\omega_r$ and reed damping $q_r$ have a strong influence on the dynamics of the system, as demonstrated by \cite{silva_interaction_2008} for the oscillation threshold. 
The complete recalibration of the model on experimental results is beyond the scope of this article.

\subsection{Comparison with experimental results from Dalmont and Frappé (2007) }

\begin{figure}[h!]
	\centering	
	\includegraphics[width=.49 \textwidth]{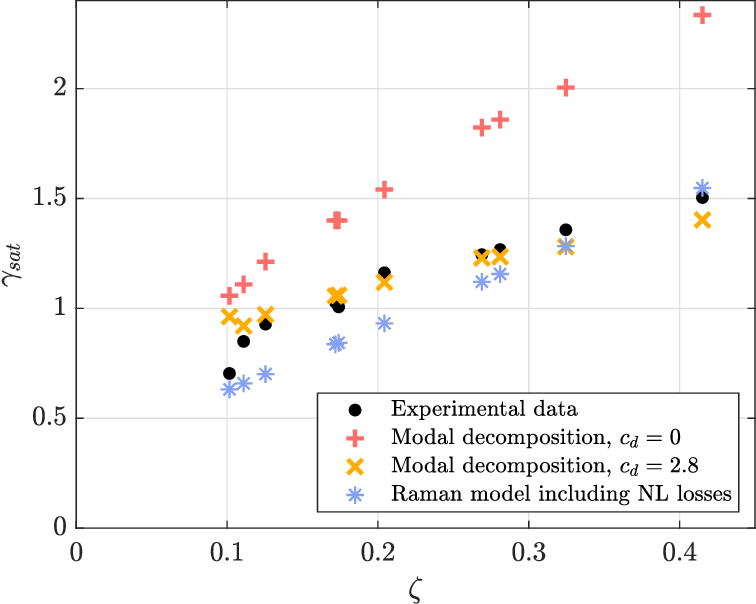}
	\caption{Experimental results from \cite{dalmont_oscillation_2007} for the saturation threshold $(\bullet)$, compared with three methods: modal decomposition without nonlinear losses $(+)$; modal decomposition including nonlinear losses $(\times)$; analytical solution provided by Raman model including nonlinear losses $(*)$.}
	\label{fig_5}
\end{figure}

An attempt to quantitatively validate the model of nonlinear losses is carried out. 
To do so, the values of the saturation threshold $\gamma_{sat}$ measured by \cite{dalmont_oscillation_2007} are employed (see Figure \ref{fig_5}). 
This threshold corresponds to the blowing pressure $\gamma$ for which $p$ is maximum during a \textit{crescendo}. 
The authors measured this threshold for different reed openings, which are transcribed here in different values of $\zeta$. 
The dimensioning pressure values $p_M$ used here correspond to the beating pressure estimated by the authors for a \textit{diminuendo}.
Furthermore, the measurements are performed on a cylindrical tube of dimensions $L=50$~cm and $R=8$~mm.
The geometry at the opening has been estimated by the authors at $c_d=2.8$. 

The experimental data are compared to simulations using the same parameters $L$ and $R$ as in the experiment. 
Simulations are performed for two cases: the first one does not take into account nonlinear losses ($c_d=0$), the other one takes them into account ($c_d=2.8$).
The simulations are performed for ascending ramps of blowing pressure varying from $\gamma=0$ to $\gamma=3$ in 10~s. 
The other parameters are the same as those given in table \ref{tab_1}. 

The model presented in this paper is also compared to the Raman model including nonlinear losses proposed by \cite{dalmont_oscillation_2007}. 
In this model, viscothermal losses are simplified by a frequency-independent coefficient $\alpha$ which replaces $\Re(\Gamma)$.
The simplicity of this model allows the authors to obtain an analytical expression of the saturation threshold $\gamma_{sat}$ as a function of the parameter $c_d$.

In the present work, the value of $\alpha$ was first adjusted at $\Re(\Gamma(j \omega_1))$, where $\omega_1$ is the {angular frequency} of the impedance peak supporting the oscillation. 
However, this value produced excessively high estimates of $\gamma_{sat}$ compared to experimental results, especially for high values of $\zeta$.
To better match the experimental data, $\alpha = 2.7\Re(\Gamma(j \omega_1)) \approx 0.147$ was chosen. 
It should be noted that the analytical expression of $\gamma_{sat}$ exhibits a linear dependence in $1/\alpha$, for low values of $\alpha$. 
Thus, $\gamma_{sat}$ is extremely sensitive to the value of $\alpha$ in Raman's model. 

Figure \ref{fig_5} illustrates the evolution of the saturation threshold as a function of $\zeta$ in the experimental case, through modal decomposition simulation, and through the analytical expression based on the Raman model.
It appears first that the saturation threshold is overestimated when nonlinear losses are not taken into account $(+)$. 
In comparison, the two models taking into account nonlinear losses produce results much closer to the experiment. 
This shows the importance of taking this phenomenon into account in wind instrument simulations.

Furthermore, for the analytical solution given by the Raman model $(*)$, the evolution of $\gamma_{sat}$ follows a straight line whose slope depends on $1/\alpha$.
Although the results are close to the experimental ones, the high sensitivity of $\gamma_{sat}$ to $\alpha$ as well as the linear evolution of $\gamma_{sat}$ reflect the excessive simplicity of the Raman model to describe the dynamics of a wind instrument.

Finally, the model presented in this paper $(\times)$ has a good agreement with the experimental data, except for very low values of the mouthpiece parameter ($\zeta<0.12$).
For these low values of $\zeta$, the model overestimates $\gamma_{sat}$ compared to the experimental results. 
This overestimation may be related to the transient behavior at extinction caused by the temporal evolution of the control parameter $\gamma$. 
A characterization of the system by continuation could give results independent of the evolution rate of $\gamma(t)$.

In conclusion, the correspondence between the results from the model presented in this article and the experimental data from \cite{dalmont_oscillation_2007} highlights its ability to describe the dynamic behavior of a simplified wind instrument at saturation. 
\section{Conclusion}

This article develops the design of a sound synthesis model of a reed instrument by modal decomposition of the input impedance, taking into account viscothermal losses as well as nonlinear losses at the end of the resonator.
The input impedance now depends on the RMS acoustic velocity at a geometric discontinuity (here, the open termination). Poles and residues resulting from the modal decomposition are fitted as a function of this velocity. In a physical model of wind instrument, the pressure-flow relation defined by the resonator is then completed by new equations which account for this dependence with the velocity at the end of the pipe.

To evaluate the ability of the model to reproduce a real phenomenon, comparisons with the experimental results of \cite{atig2004saturation} and \cite{dalmont_oscillation_2007} have been made. In the first case, simulations show a qualitatively similar behavior regarding the evolution of the extinction threshold depending on the geometry at the open end. In the second case, the model gives a good correspondence with the experimental results, in particular compared to a model without nonlinear losses.

This formalism could be promising in the sound synthesis of wind instruments in real time, which employs the modal decomposition formalism in order to keep a minimum of memory. The inclusion of nonlinear losses in a complete clarinet model could more accurately translate dynamic phenomena essential to the experience of the musician, as suggested by \cite{dalmont_oscillation_2007} : "any realistic model of the clarinet should include nonlinear losses in the side holes".

\section*{Acknowledgments}
This study has been supported by the French ANR LabCom LIAMFI (ANR-16-LCV2-007-01). 
The authors thank S. Maugeais and J.-P. Dalmont for their valuable comments.

\bibliography{biblio}
\newpage 
\listoffigures
\listoftables
\end{document}